\documentclass[aps,prl,twocolumn,groupedaddress,showpacs]{revtex4-1}
\usepackage{enumerate}
\usepackage{amsmath, amsthm, amssymb}
\usepackage{color,calc,graphicx,subfig}

\newcommand{\ket}[1]{| #1 \rangle}
\newcommand{\bra}[1]{\langle #1 |}
\newcommand{\ip}[2]{\langle #1 | #2 \rangle}

\newtheorem{thm}{Proposition}
\newtheorem{lemma}{Lemma}
\newtheorem{corollary}{Corollary}

\begin{document}


\title{Nonlocality of Symmetric States}

\author{Zizhu Wang}
\email[]{zizhu.wang@telecom-paristech.fr}

\author{Damian Markham}
\email[]{damian.markham@telecom-paristech.fr}
\affiliation{CNRS LTCI, D\'{e}partement Informatique et R\'{e}seaux, Telecom ParisTech, 23 avenue d'Italie, CS 51327,  75214 Paris CEDEX 13, France}

\begin{abstract}

In this paper we study the nonlocal properties of permutation symmetric states of $n$-qubits.
We show that all these states are nonlocal, via an extended version of the Hardy paradox and associated inequalities.
Natural extensions of both the paradoxes and the inequalities are developed which relate different entanglement classes to different nonlocal features.
Belonging to a given entanglement class will guarantee the violation of associated Bell inequalities which see the persistence of correlations to subsets of players, whereas there are states outside that class which do not violate.
\end{abstract}


\maketitle

\textit{Introduction.}---
Nonlocality is a foundational feature of quantum mechanics and is increasingly becoming recognised as a key resource for quantum information theory, for example in device independence~\cite{PhysRevLett.95.010503,colbeck2009quantum,Pironio2010fk,masanes2011secure}, communication complexity~\cite{RevModPhys.82.665} and measurement based quantum computation  ~\cite{PhysRevLett.86.5188,PhysRevLett.102.050502}.
Related, though not equivalent is the feature of entanglement. In the multipartite setting entanglement is very complicated, different {\it classes} of entanglement exist, each having potentially different roles as resources. Very little is currently known if, and how, the richness of multipartite entanglement is reflected in nonlocal features, with some very recent breakthroughs~\cite{PhysRevLett.108.110501}.

We explore the nonlocal features of permutation symmetric states of qubits. This set of states are useful in a variety of quantum information tasks, they occur naturally as ground states in some Bose-Hubbard models, and are amongst the most developed experimentally. Relatively little is known about the nonlocality of permutation symmetric states, mostly restricted to W and GHZ states~\cite{mermin1995best,PhysRevA.65.032108,1367-2630-13-5-053054}.

Knowing more about their nonlocality would help understand more their potential as resources for quantum information processing, and understand better the relationship between the subtleties of multipartite entangled states and nonlocal features.
Recently we begin to get a better understanding of the entanglement features of symmetric states using the Majorana representation \cite{majorana1932atomi,PhysRevLett.103.070503,aulbach2011symmetric,PhysRevLett.106.180502,PhysRevA.83.042332,aulbach2010maximally}. Here we use the same tool to study the states' nonlocality, allowing us to compare it to entanglement easily.

Consider $n$ parties, indexed by $i$, each of which make a measurent in a chosen setting $M_i$, and get result $r_i$. We will consider a choice of two settings, each with two outcomes. A probability distribution over the measurements is \emph{local} or admits a \emph{local hidden varible (LHV)} description if the joint probability distribution can be written as the product of individual probabilities given the value of some hidden variable $\lambda$:
\begin{align}
P(r_1,\ldots,r_n|M_1,\ldots,M_n)=\int \rho(\lambda)\prod_i P(r_i|M_i,\lambda) d\lambda\label{lhv},
\end{align}
where $P(r_i|M_i,\lambda)$ is the probability for the $i$th party to obtain the result $r_i$ when using the measurement setting $M_i$ while having $\lambda$ as the value of the hidden variable. $\rho(\lambda)$ is the probability distribution of $\lambda$. $P(r_1,\ldots,r_n|M_1,\ldots,M_n)$ is the joint probability distribution when all $n$ parties measure using the settings $M_1,\ldots,M_n$ and obtain the results $r_1,\ldots,r_n$. Often we will ignore the lower index when position is obvious. It is obvious that local measurement on any separable state admits an LHV description. However, nonlocality does not follow directly from entanglement~\cite{RevModPhys.81.865}.

The Hardy paradox has been proposed as an ``almost probability-free'' test of the nonlocality of almost all bipartite entangled quantum states~\cite{PhysRevLett.71.1665,PhysRevLett.73.2279}. We first show that all permutation symmetric states of $n$ qubits can violate an extended version of the Hardy Paradox and associated inequalities. While there exists earlier work generalizing the Hardy paradox to $n$-party~\cite{ghosh2010chainjournal} equivalent to ours, we give a constructive way of finding measurements needed by the $n$-party paradox to show that all permutation symmetric states are nonlocal.

\textit{The Hardy paradox and inequality for all permutation symmetric states.}---The original Hardy paradox consists of four probabilistic conditions that we impose on the outcomes of an experiment involving two parties~\cite{PhysRevLett.71.1665}~\cite{PhysRevLett.73.2279}. These conditions are individually compatible with the definition of a hidden variable theory given in (\ref{lhv}). But when taken together, they lead to a logical contradiction.  Hardy showed that for almost all bipartite entangled states, there exist measurement settings to satisfy all these conditions, thus showing the incompatibility of LHV theory and quantum mechanics. The only exception are the maximally entangled states. Fortunately, the nonlocality of the maximally entangled states was proven before~\cite{bell1964einstein}~\cite{PhysRevLett.23.880}.

A multiparty extension of the Hardy paradox can be constructed as follows: first suppose there are $n$ players involved in an experiment. Each player can choose to measure one of two possible measurement settings labeled $0$ or $1$, and get one of two possible outcomes, also labeled $0$ or $1$. The first probabilistic condition we impose is that if everyone measures in the $0$ basis, then sometimes everyone gets the result $0$:
\begin{align}
P(00\ldots00|00\ldots00)>0.\label{p1}
\end{align}

The next $n$ conditions are the same as above for $n-1$ players, but now if one player measures in the setting $1$ instead of the setting $0$  they will never get the result $0$.
\begin{align}
P(00\ldots00&|\pi(00\ldots01))=0,\label{p2}
\end{align}
for all permutations $\pi$ of bit strings with one $1$ and $n-1$ zeros. Let us consider the implications imposed if these arise from an LHV model. We see that~(\ref{lhv}) and~(\ref{p1}) imply that there exists at least one value of $\lambda$ such that $\forall i, P(0_i|0_i \lambda)>0$. Then, for this particular value of $\lambda$, we know from~(\ref{p2}) that $\forall i, P(0_i|1_i \lambda)=0$. Since there are only two possible outcomes for each measurement setting, (\ref{p2}) imply that for this value of $\lambda$, should everyone instead chose to measure in setting $1$, they must all get result $1$ with certainty.

The last condition we impose contradicts the conclusion we get above. If everyone measures in setting $1$, then they will never all get the result $1$:
\begin{align}
P(11\ldots1|11\ldots11)=0.\label{p5}
\end{align}
Clearly (\ref{p1}), (\ref{p2}) and (\ref{p5}) are not possible within LHV. Note that in the case where $n=2$, we recover the original Hardy paradox~\cite{PhysRevLett.71.1665},~\cite{PhysRevLett.73.2279}.

However, we will now give a constructive procedure to find the bases $0$ and $1$ for almost all permutation symmetric states such that the conditions~(\ref{p1}) to~(\ref{p5}) are all satisfied. As a prerequisite, let us recall some basic properties of permutation symmetric states and the Majorana representation. More details can be found in the Supplemental Material~\cite{appendices} and in~\cite{majorana1932atomi}~\cite{PhysRevA.83.042332}~\cite{aulbach2010maximally}~\cite{penrose20001}~\cite{martinthesis}.

A permutation symmetric state of $n$ qubits can be written in the form $\ket{\psi}=\sum_{k=0}^{n} c_k\ket{S(n,k)}$, where $\ket{S(n,k)}={n\choose k}^{-\frac{1}{2}}\sum_{perm}\ket{\underbrace{0\ldots0}_{n-k}\underbrace{1\ldots1}_{k}}$ are Dicke states.

In the Majorana representation, the state $\ket{\psi}$ is written as a sum of permutations of the tensor product of $n$ qubits $\{\ket{\eta_1}\ldots\ket{\eta_n}\}$, called the \emph{Majorana Points (MPs)} of the state $\ket{\psi}$:
\begin{align}
\ket{\psi}=K\sum_{perm}\ket{\eta_1\ldots\eta_n}\label{decomp}.
\end{align}
$K$ is a normalization constant which depends on the overlap between different MPs. In the Majorana representation, local unitaries of the form $U^{\otimes n}$ simply rotates all Majorana points at the same time, thus equivalent to a rotation of the Bloch Sphere.

Permutation symmetry also persists to subspaces. If $\ket{\psi}$ is a permutation symmetric state of $n$ qubits, then for any single qubit state $\ket{\chi}$ (ignoring normalisation), the $(n-1)$-qubit state $\ip{\chi}{\psi}$ is also permutation symmetric:
\begin{align}
\ip{\chi}{\psi}=\sum_{i=1}^{n} C_{i}\sum_{perm}\ket{\underbrace{\eta_1\ldots\eta_n}_{\{1,\ldots,n\}\setminus i}},\label{projection}
\end{align}
 where $C_i=\ip{\chi}{\eta_i}$ and $\{1,\ldots,n\}\setminus i$ means that we discard the MP $\ket{\eta_i}$.

The equation below holds if and only if  $\ket{\eta_i}$ is an MP of  $\ket{\psi}$, and  $\ket{\eta_i^{\perp}}$ is its antipodal point on the Bloch sphere:
\begin{align}
(\bra{\eta_i^{\perp}})^{\otimes n}\ket{\psi}=0.\label{orthogonal}
\end{align}

We will now see how to choose the measurement bases that satisfy (\ref{p1}) - (\ref{p5}) for almost all permutation symmetric states. First of all, (\ref{orthogonal}) can be seen as the probability amplitude that gives~(\ref{p5}) if we restrict the measurement to be projective and take the $\{\ket{\eta_i},\ket{\eta_i^{\perp}}\}$ basis as measurement setting $1$ for all parties.

For condition~(\ref{p2}), if one can be satisfied, then by symmetry of the state the rest are satisfied automatically. Consider the projection of the state $\ket{\psi}$ on one of its MPs $\ket{\eta_i}$. By~(\ref{projection}) this gives us a new permutation symmetric state of $(n-1)$ qubits:
\begin{align}
\ket{\psi'}=\ip{\eta_i}{\psi}=\sum_{j=1}^{n} C_{j}\sum_{perm}\ket{\underbrace{\eta_1\ldots\eta_n}_{\{1,\ldots,n\}\setminus j}},\label{projection2}
\end{align}
 with $C_j=\ip{\eta_i}{\eta_j}$.
The state $\ket{\psi'}$ has $(n-1)$ MPs, possibly different from the MPs of $\ket{\psi}$. In fact, the proposition below shows that for all permutation symmetric states except Dicke states, there is always at least one MP of $\ket{\psi'}$ that is different from all the MPs of $\ket{\psi}$.
\begin{thm}\label{allbutdicke}
Let $S_{\psi}:=\{\ket{\eta_1},\ket{\eta_2},\ldots,\ket{\eta_n}\}$ be the set of MPs of the state $\ket{\psi}$. Let $S_{\psi_i}:=\{\ket{\mu_1},\ket{\mu_2},\ldots,\ket{\mu_{n-1}}\}$ be the set of MPs of the state $\ket{\psi_i}=\ip{\eta_i}{\psi}$. Then $S_{\psi_i}\subseteq S_{\psi}$ iff $\ket{\psi}$ is a Dicke state up to rotations of the Bloch Sphere.
\end{thm}
See Supplemental Material~\cite{appendices} for proof.

Let $\ket{\mu_i}$ be an MP of the state $\ket{\psi'}$ as defined in (\ref{projection2}) that is different from all the MPs of $\ket{\psi}$, then by~(\ref{orthogonal})
\begin{align}
(\bra{\mu_i^{\perp}})^{\otimes n-1}\ket{\psi'}=\ip{\eta_i\underbrace{\mu_i^{\perp}\ldots\mu_i^{\perp}}_{n-1}}{\psi}=0.\label{amp2}
\end{align}

By choosing basis $\{\ket{\mu_i},\ket{\mu_i^{\perp}}\}$ as measurement setting $0$, for all parties, the probability amplitude~(\ref{amp2}) implies the satisfaction of condition~(\ref{p2}) by symmetry. Because $\ket{\mu_i}$ is not an MP of $\ket{\psi}$, $\ip{\underbrace{\mu_i^{\perp}\ldots\mu_i^{\perp}}_{n}}{\psi}\neq 0$. Thus~(\ref{p1}) is satisfied automatically also. By Proposition~\ref{allbutdicke}, this procedure of choosing measurement settings $0$ and $1$ works for all permutation symmetric states except Dicke states.

The paradox itself, however, can not be tested directly by experiments because in real experiments, when taking real-world noise and inaccuracy into account, we will never see probabilities getting exactly zero. To make the paradox more noise tolerant, we make it into an inequality. The LHV upper bound of the inequality can be violated by the amount of~(\ref{p1}) when use the procedure given above to perform a quantum experiment.

\begin{thm}\label{ineq}
The Bell operator for $n$ systems
\begin{align*}
\mathcal{P}^n:= &P(0\ldots0|00\ldots00)\\
&-\sum_{\pi}P(00\ldots00|\pi(00\ldots01))\\
&-P(1\ldots1|11\ldots11)
\end{align*}
is bounded under LHV as $\mathcal{P}^n \leq 0$.
\end{thm}
See Supplemental Material~\cite{appendices} for proof (See also~\cite{ghosh2010chainjournal} for an alternative proof).

Although the procedure we used to find measurement settings does not work for Dicke states, the inequality above can be violated by Dicke states in a $2$ settings/$2$ outcomes experiment:
\begin{thm}\label{alldicke}
There exists an angle $0<\theta<\pi$ such that all Dicke states $\ket{S(n,k)}$ ($\{k,n\}\in \mathbb{N}, 1<k<n$) violate the inequality in Proposition~\ref{ineq} when using the measurement setting $\{\ket{+}, \ket{-}\}$ as setting $0$ and $\{\cos{\frac{\theta}{2}}\ket{0}-\sin{\frac{\theta}{2}}\ket{1}, \sin{\frac{\theta}{2}}\ket{0}+\cos{\frac{\theta}{2}}\ket{1}\}$ as setting $1$.
\end{thm}
See Supplemental Material~\cite{appendices} for proof.

\textit{Entanglement classes and nonlocality.}---In the Majorana representation of a permutation symmetric state~(\ref{decomp}), the Majorana Points $\{\ket{\eta_1},\ldots,\ket{\eta_n}\}$ are not necessarily all distinct. In this section we will slightly alter our notation to incorporate the notion of multiplicity or degeneracy, which means several MPs are ``sitting on top of one another''. In the new notation, we use $d_i$ to denote the degeneracy of the MP $\ket{\eta_i}$. So (\ref{decomp}) becomes
\begin{align}
\ket{\psi}=K\sum_{perm}\ket{\eta_1^{d_1}\eta_2^{d_2}\ldots\eta_l^{d_l}},\label{multdecomp}\\
\forall i\neq j, \ket{\eta_i}\neq\ket{\eta_j}, \sum_{i=1}^l d_i=n.\nonumber
\end{align}
Also, (\ref{orthogonal}) becomes
\begin{align}
(\bra{\eta_i^{\perp}})^{\otimes k}\ket{\psi}=0.\label{multi_ortho}
\end{align}
where $(n-d_i)<k\leq n$. Degeneracy of points cannot change under local operations and classical communication, even stochastically (SLOCC)~\cite{PhysRevLett.103.070503} (see also~\cite{PhysRevLett.106.180502,aulbach2011symmetric}). Thus different degeneracies of points, corresponds to different entanglement classes.

Taking degeneracy into account, we can extend the paradox by considering subsets of players. Translating (\ref{multi_ortho}) to statements of probabilities, we see that the correlations of  (\ref{p5}) persist to fewer players
\begin{align}
P(\underbrace{11\ldots1}_{k}&|\underbrace{11\ldots1}_{k})=0,\label{mp5}
\end{align}
for $(n-d_i)<k\leq n$.
The inequality in Proposition~\ref{ineq} can be extended naturally to:
\begin{align} \label{def: degQ}
\mathcal{Q}_{d}^n := &\mathcal{P}^n - P(\underbrace{11\ldots1}_{n-1}|\underbrace{11\ldots1}_{n-1}) - ... - P(\underbrace{11\ldots1}_{n-d+1}|\underbrace{11\ldots1}_{n-d+1})\nonumber\\
& \leq 0.
\end{align}
The LHV upper bound holds because $\mathcal{P}^n$ is negative by Proposition~\ref{ineq}, and we are only subtracting positive probabilities from it.

\begin{figure}[ht]
  \centering
  \subfloat[$Q_3^4\leq -0.0609$]{\label{ex:t}\includegraphics[width=120px,keepaspectratio=true]{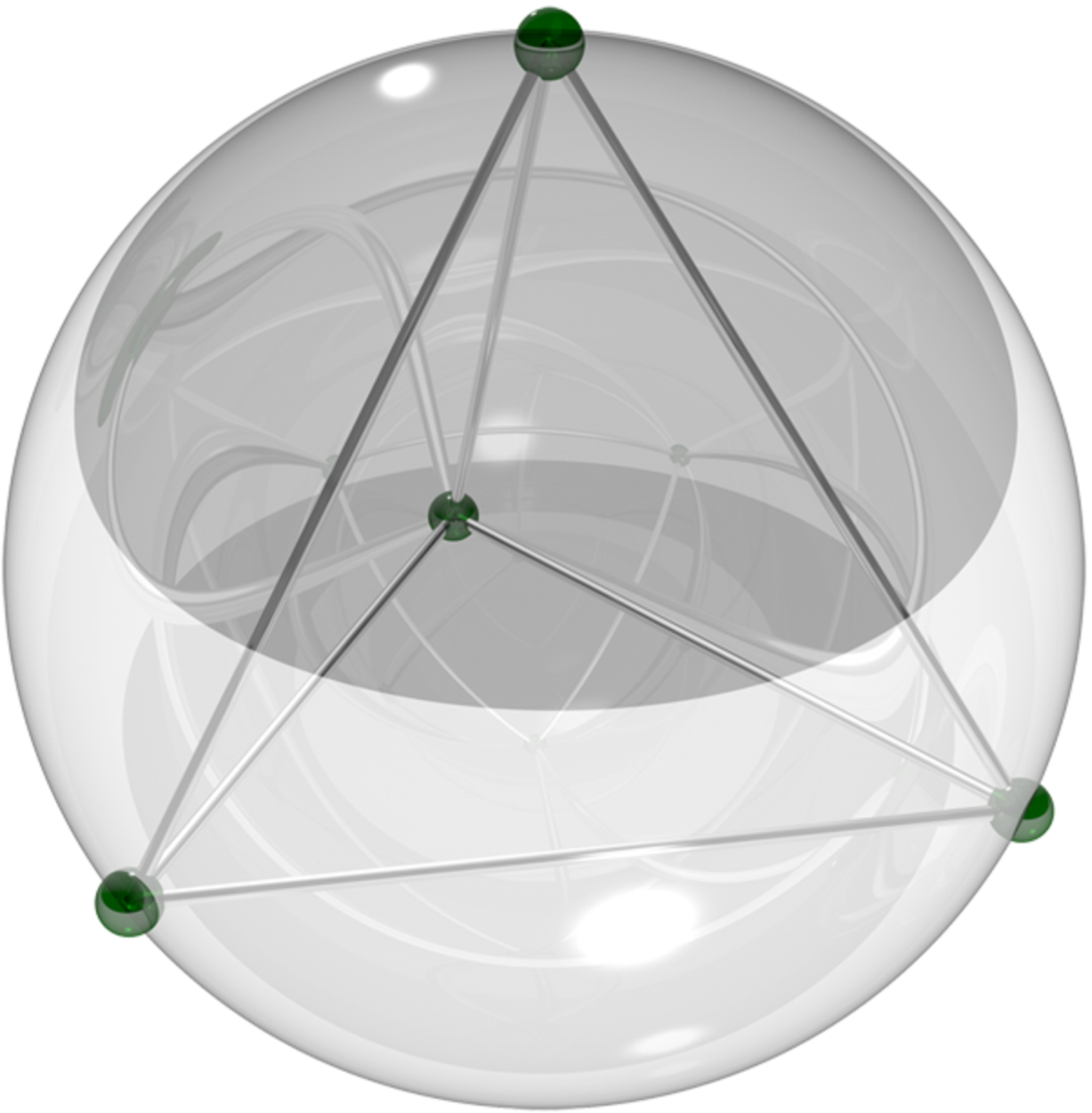}}
  \,
  \subfloat[$Q_3^4\geq 0.0141$]{\label{ex:000+}\includegraphics[width=120px,keepaspectratio=true]{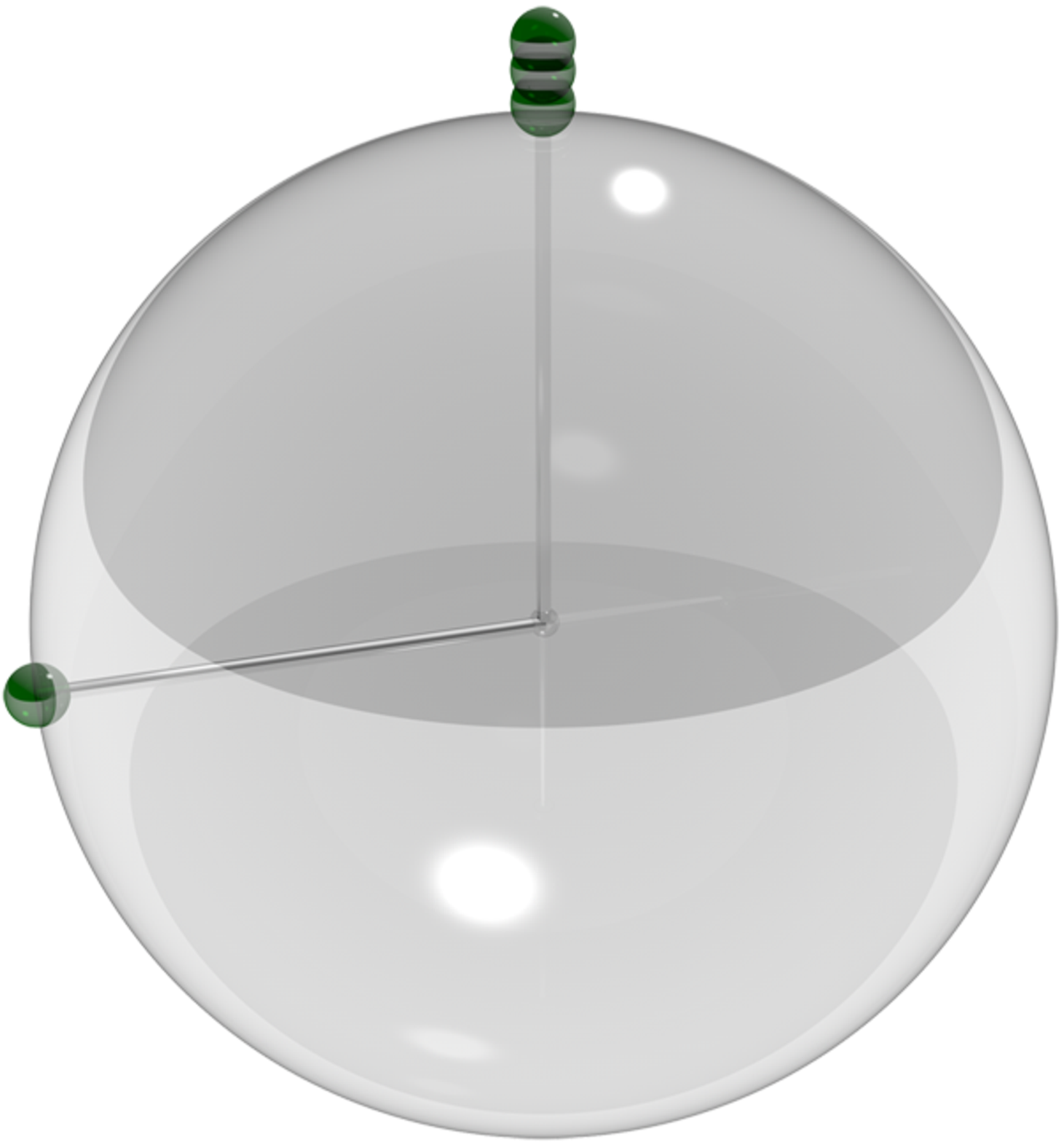}}
  \caption{The state a) $\ket{T}=\sqrt{\frac{1}{3}}\ket{0000}+\sqrt{\frac{2}{3}}\ket{S(4,3)}$ does not violate $Q_3^4$ while all states with degeneracy $d=3$ do, such as the state b) $\ket{D_3}=K\sum_{perm}\ket{000+}$.}
  \label{examples}
\end{figure}

We can now see how this inequality allows us to differentiate different entanglement classes  via degeneracy classes. First, it is clear that all states with at least one MP with degeneracy $d$ will be able to violate inequality (\ref{def: degQ}). Second, this is not true for all states with lower degeneracy. Note that we cannot hope that all states with maximum degeneracy less than $d$ do not violate $Q_d^n$, or indeed any inequality violated by all states with degeneracy $d$, since we can lower the degeneracy by moving one MP away by an arbitrary distance. In this sense the best that we could hope for is that certain states, or classes of states outside the associated entanglement class cannot violate. It can be checked by using semidefinite programming techniques similar to the ones used in~\cite{PhysRevLett.108.110501} that the tetrahedron state, shown in Fig.~\ref{ex:t}) does not violate $Q_3^4$ while the state in Fig.~\ref{ex:000+}) does. A similar separation between W states and Schmidt states (states such that removal of one system destroys the entanglement) has been found recently in~\cite{PhysRevLett.108.110501}.

Furthermore, if all parties measure projectively in the same basis, the only way they can satisfy conditions (3), (4), (5) as well as (\ref{mp5}) is if they have at least one MP with degeneracy $d \geq d_i$. This is because (5) implies the basis is an MP, and (\ref{mp5}) bounds the degeneracy of it. In this sense one can witness different entanglement classes. One may further expect that with the same measurement restrictions high violation of the inequalities depends on the degeneracies. In this way these extensions probe the different entanglement types given be differing degenerecies~\cite{PhysRevLett.103.070503}~\footnote{Even though it would be useful in an LHV test to also test entanglement class in a device independent way, it does not make sense to assert that all parties `measure in the same basis'. One may append to the list of conditions (\ref{p1})-(\ref{p5}) some extra conditions which effectively imply the state is symmetric with respect to the basis given by measurement setting $1$. However, whilst one may be able to define Hardy type paradoxes which can be used to identify classes of entanglement in this way, these conditions would be difficult to fit into an inequality, and further it is not clear that it would be possible at all that such inequalities would also strictly separate degeneracy classes.}.

\textit{Conclusions.}---In this paper we have presented new Hardy type paradoxes and associated Bell inequalities, and given a procedure to find bases to show violation for all permutation symmetric states of qubits, which can be understood as the generalization of Gisin's Theorem~\cite{gisin1991bell} to permutation symmetric states. One property of the inequalities which is obvious immediately is that the inequalities are written in terms of probabilities and cannot be extended to normal correlation operators alone. Since the number of settings and outcomes is two, and since it also works for all extended GHZ states, it provides an example of a Bell operator which is more powerful than possible by correlation operators alone~\cite{PhysRevLett.88.210402}.

The structure of nonlocal features is also explored via these methods. Natural extensions of both the paradox and the inequalities are presented which relate to different entanglement classes (specifically, degeneracy classes \cite{PhysRevLett.103.070503}). On the one hand this provides a witness to discriminate different entanglement classes if measurements are set as outlined. On the other hand, states of minimum degeneracy can certainly violate associated inequalities, whereas other states will not, for example the state $|T\rangle$ as shown here, no matter what measurements are made, hence providing a possibility for device independent testing of state class, as was done in~\cite{PhysRevLett.108.110501} for W and Schmidt state classes. Thus high degeneracy of MPs, originally considered in terms of the abstract definition of SLOCC classification, has a practical application demonstrating the persistence of nonlocal correlations to fewer systems. In this sense, the difference noted between W and GHZ states in~\cite{PhysRevA.65.032108} are just examples this more general feature.

As nonlocal features become more and more recognised as important for applications in quantum information we can expect that these results will lead to better understanding of the usefulness of permutation symmetric states. We also note that preparation of these states, and indeed the projective measurements presented is well within experimental grasp in several different possible experimental frameworks~\cite{PhysRevLett.103.020503,PhysRevLett.103.020504,PhysRevLett.102.053601}.

We thank {\v C}aslav Brukner, Sudha Shenoy and Adel Sohbi for helpful comments. This work was supported by the joint ANR-NSERC grant ``Fundamental Research in Quantum Networks and Cryptography (FREQUENCY)''.

\bibliography{biblio}

\begin{thebibliography}{32}%
\makeatletter
\providecommand \@ifxundefined [1]{%
 \@ifx{#1\undefined}
}%
\providecommand \@ifnum [1]{%
 \ifnum #1\expandafter \@firstoftwo
 \else \expandafter \@secondoftwo
 \fi
}%
\providecommand \@ifx [1]{%
 \ifx #1\expandafter \@firstoftwo
 \else \expandafter \@secondoftwo
 \fi
}%
\providecommand \natexlab [1]{#1}%
\providecommand \enquote  [1]{``#1''}%
\providecommand \bibnamefont  [1]{#1}%
\providecommand \bibfnamefont [1]{#1}%
\providecommand \citenamefont [1]{#1}%
\providecommand \href@noop [0]{\@secondoftwo}%
\providecommand \href [0]{\begingroup \@sanitize@url \@href}%
\providecommand \@href[1]{\@@startlink{#1}\@@href}%
\providecommand \@@href[1]{\endgroup#1\@@endlink}%
\providecommand \@sanitize@url [0]{\catcode `\\12\catcode `\$12\catcode
  `\&12\catcode `\#12\catcode `\^12\catcode `\_12\catcode `\%12\relax}%
\providecommand \@@startlink[1]{}%
\providecommand \@@endlink[0]{}%
\providecommand \url  [0]{\begingroup\@sanitize@url \@url }%
\providecommand \@url [1]{\endgroup\@href {#1}{\urlprefix }}%
\providecommand \urlprefix  [0]{URL }%
\providecommand \Eprint [0]{\href }%
\@ifxundefined \urlstyle {%
  \providecommand \doi  [0]{\begingroup \@sanitize@url \@doi}%
  \providecommand \@doi [1]{\endgroup \@@startlink {\doibase
  #1}doi:\discretionary {}{}{}#1\@@endlink }%
}{%
  \providecommand \doi  [0]{doi:\discretionary{}{}{}\begingroup
  \urlstyle{rm}\Url }%
}%
\providecommand \doibase [0]{http://dx.doi.org/}%
\providecommand \Doi [0]{\begingroup \@sanitize@url \@Doi }%
\providecommand \@Doi  [1]{\endgroup\@@startlink{\doibase#1}\@@Doi}%
\providecommand \@@Doi [1]{#1\@@endlink}%
\providecommand \selectlanguage [0]{\@gobble}%
\providecommand \bibinfo  [0]{\@secondoftwo}%
\providecommand \bibfield  [0]{\@secondoftwo}%
\providecommand \translation [1]{[#1]}%
\providecommand \BibitemOpen [0]{}%
\providecommand \bibitemStop [0]{}%
\providecommand \bibitemNoStop [0]{.\EOS\space}%
\providecommand \EOS [0]{\spacefactor3000\relax}%
\providecommand \BibitemShut  [1]{\csname bibitem#1\endcsname}%
\bibitem [{\citenamefont {Barrett}\ \emph {et~al.}(2005)\citenamefont
  {Barrett}, \citenamefont {Hardy},\ and\ \citenamefont
  {Kent}}]{PhysRevLett.95.010503}%
  \BibitemOpen
  \bibfield  {author} {\bibinfo {author} {\bibfnamefont {J.}~\bibnamefont
  {Barrett}}, \bibinfo {author} {\bibfnamefont {L.}~\bibnamefont {Hardy}}, \
  and\ \bibinfo {author} {\bibfnamefont {A.}~\bibnamefont {Kent}},\ }\Doi
  {10.1103/PhysRevLett.95.010503} {\bibfield  {journal} {\bibinfo  {journal}
  {Phys. Rev. Lett.},\ }\textbf {\bibinfo {volume} {95}},\ \bibinfo {pages}
  {010503} (\bibinfo {year} {2005})}\BibitemShut {NoStop}%
\bibitem [{\citenamefont {Colbeck}(2006)}]{colbeck2009quantum}%
  \BibitemOpen
  \bibfield  {author} {\bibinfo {author} {\bibfnamefont {R.}~\bibnamefont
  {Colbeck}},\ }\emph {\bibinfo {title} {Quantum and relativistic protocols for
  secure multi-party computation}},\ \href@noop {} {Ph.D. thesis},\ \bibinfo
  {school} {University of Cambridge} (\bibinfo {year} {2006})\BibitemShut
  {NoStop}%
\bibitem [{\citenamefont {Pironio}\ \emph {et~al.}(2010)\citenamefont
  {Pironio}, \citenamefont {Ac{\'\i}n}, \citenamefont {Massar} \emph
  {et~al.}}]{Pironio2010fk}%
  \BibitemOpen
  \bibfield  {author} {\bibinfo {author} {\bibfnamefont {S.}~\bibnamefont
  {Pironio}}, \bibinfo {author} {\bibfnamefont {A.}~\bibnamefont {Ac{\'\i}n}},
  \bibinfo {author} {\bibfnamefont {S.}~\bibnamefont {Massar}},  \emph
  {et~al.},\ }\href {http://dx.doi.org/10.1038/nature09008} {\bibfield
  {journal} {\bibinfo  {journal} {Nature},\ }\textbf {\bibinfo {volume}
  {464}},\ \bibinfo {pages} {1021} (\bibinfo {year} {2010})}\BibitemShut
  {NoStop}%
\bibitem [{\citenamefont {Masanes}\ \emph {et~al.}(2011)\citenamefont
  {Masanes}, \citenamefont {Pironio},\ and\ \citenamefont
  {Acin}}]{masanes2011secure}%
  \BibitemOpen
  \bibfield  {author} {\bibinfo {author} {\bibfnamefont {L.}~\bibnamefont
  {Masanes}}, \bibinfo {author} {\bibfnamefont {S.}~\bibnamefont {Pironio}}, \
  and\ \bibinfo {author} {\bibfnamefont {A.}~\bibnamefont {Acin}},\ }\href@noop
  {} {\bibfield  {journal} {\bibinfo  {journal} {Nature Communications},\
  }\textbf {\bibinfo {volume} {2}},\ \bibinfo {pages} {238} (\bibinfo {year}
  {2011})}\BibitemShut {NoStop}%
\bibitem [{\citenamefont {Buhrman}\ \emph {et~al.}(2010)\citenamefont
  {Buhrman}, \citenamefont {Cleve}, \citenamefont {Massar},\ and\ \citenamefont
  {de~Wolf}}]{RevModPhys.82.665}%
  \BibitemOpen
  \bibfield  {author} {\bibinfo {author} {\bibfnamefont {H.}~\bibnamefont
  {Buhrman}}, \bibinfo {author} {\bibfnamefont {R.}~\bibnamefont {Cleve}},
  \bibinfo {author} {\bibfnamefont {S.}~\bibnamefont {Massar}}, \ and\ \bibinfo
  {author} {\bibfnamefont {R.}~\bibnamefont {de~Wolf}},\ }\Doi
  {10.1103/RevModPhys.82.665} {\bibfield  {journal} {\bibinfo  {journal} {Rev.
  Mod. Phys.},\ }\textbf {\bibinfo {volume} {82}},\ \bibinfo {pages} {665}
  (\bibinfo {year} {2010})}\BibitemShut {NoStop}%
\bibitem [{\citenamefont {Raussendorf}\ and\ \citenamefont
  {Briegel}(2001)}]{PhysRevLett.86.5188}%
  \BibitemOpen
  \bibfield  {author} {\bibinfo {author} {\bibfnamefont {R.}~\bibnamefont
  {Raussendorf}}\ and\ \bibinfo {author} {\bibfnamefont {H.~J.}\ \bibnamefont
  {Briegel}},\ }\Doi {10.1103/PhysRevLett.86.5188} {\bibfield  {journal}
  {\bibinfo  {journal} {Phys. Rev. Lett.},\ }\textbf {\bibinfo {volume} {86}},\
  \bibinfo {pages} {5188} (\bibinfo {year} {2001})}\BibitemShut {NoStop}%
\bibitem [{\citenamefont {Anders}\ and\ \citenamefont
  {Browne}(2009)}]{PhysRevLett.102.050502}%
  \BibitemOpen
  \bibfield  {author} {\bibinfo {author} {\bibfnamefont {J.}~\bibnamefont
  {Anders}}\ and\ \bibinfo {author} {\bibfnamefont {D.~E.}\ \bibnamefont
  {Browne}},\ }\Doi {10.1103/PhysRevLett.102.050502} {\bibfield  {journal}
  {\bibinfo  {journal} {Phys. Rev. Lett.},\ }\textbf {\bibinfo {volume}
  {102}},\ \bibinfo {pages} {050502} (\bibinfo {year} {2009})}\BibitemShut
  {NoStop}%
\bibitem [{\citenamefont {Brunner}\ \emph {et~al.}(2012)\citenamefont
  {Brunner}, \citenamefont {Sharam},\ and\ \citenamefont
  {V\'ertesi}}]{PhysRevLett.108.110501}%
  \BibitemOpen
  \bibfield  {author} {\bibinfo {author} {\bibfnamefont {N.}~\bibnamefont
  {Brunner}}, \bibinfo {author} {\bibfnamefont {J.}~\bibnamefont {Sharam}}, \
  and\ \bibinfo {author} {\bibfnamefont {T.}~\bibnamefont {V\'ertesi}},\ }\Doi
  {10.1103/PhysRevLett.108.110501} {\bibfield  {journal} {\bibinfo  {journal}
  {Phys. Rev. Lett.},\ }\textbf {\bibinfo {volume} {108}},\ \bibinfo {pages}
  {110501} (\bibinfo {year} {2012})}\BibitemShut {NoStop}%
\bibitem [{\citenamefont {Mermin}(1995)}]{mermin1995best}%
  \BibitemOpen
  \bibfield  {author} {\bibinfo {author} {\bibfnamefont {N.}~\bibnamefont
  {Mermin}},\ }\href@noop {} {\bibfield  {journal} {\bibinfo  {journal} {Annals
  of the New York Academy of Sciences},\ }\textbf {\bibinfo {volume} {755}},\
  \bibinfo {pages} {616} (\bibinfo {year} {1995})}\BibitemShut {NoStop}%
\bibitem [{\citenamefont {Cabello}(2002)}]{PhysRevA.65.032108}%
  \BibitemOpen
  \bibfield  {author} {\bibinfo {author} {\bibfnamefont {A.}~\bibnamefont
  {Cabello}},\ }\Doi {10.1103/PhysRevA.65.032108} {\bibfield  {journal}
  {\bibinfo  {journal} {Phys. Rev. A},\ }\textbf {\bibinfo {volume} {65}},\
  \bibinfo {pages} {032108} (\bibinfo {year} {2002})}\BibitemShut {NoStop}%
\bibitem [{\citenamefont {Heaney}\ \emph {et~al.}(2011)\citenamefont {Heaney},
  \citenamefont {Cabello}, \citenamefont {Santos},\ and\ \citenamefont
  {Vedral}}]{1367-2630-13-5-053054}%
  \BibitemOpen
  \bibfield  {author} {\bibinfo {author} {\bibfnamefont {L.}~\bibnamefont
  {Heaney}}, \bibinfo {author} {\bibfnamefont {A.}~\bibnamefont {Cabello}},
  \bibinfo {author} {\bibfnamefont {M.~F.}\ \bibnamefont {Santos}}, \ and\
  \bibinfo {author} {\bibfnamefont {V.}~\bibnamefont {Vedral}},\ }\href
  {http://stacks.iop.org/1367-2630/13/i=5/a=053054} {\bibfield  {journal}
  {\bibinfo  {journal} {New Journal of Physics},\ }\textbf {\bibinfo {volume}
  {13}},\ \bibinfo {pages} {053054} (\bibinfo {year} {2011})}\BibitemShut
  {NoStop}%
\bibitem [{\citenamefont {Majorana}(1932)}]{majorana1932atomi}%
  \BibitemOpen
  \bibfield  {author} {\bibinfo {author} {\bibfnamefont {E.}~\bibnamefont
  {Majorana}},\ }\href@noop {} {\bibfield  {journal} {\bibinfo  {journal} {Il
  Nuovo Cimento},\ }\textbf {\bibinfo {volume} {9}},\ \bibinfo {pages} {43}
  (\bibinfo {year} {1932})}\BibitemShut {NoStop}%
\bibitem [{\citenamefont {Bastin}\ \emph
  {et~al.}(2009){\natexlab{a}}\citenamefont {Bastin}, \citenamefont {Krins},
  \citenamefont {Mathonet} \emph {et~al.}}]{PhysRevLett.103.070503}%
  \BibitemOpen
  \bibfield  {author} {\bibinfo {author} {\bibfnamefont {T.}~\bibnamefont
  {Bastin}}, \bibinfo {author} {\bibfnamefont {S.}~\bibnamefont {Krins}},
  \bibinfo {author} {\bibfnamefont {P.}~\bibnamefont {Mathonet}},  \emph
  {et~al.},\ }\Doi {10.1103/PhysRevLett.103.070503} {\bibfield  {journal}
  {\bibinfo  {journal} {Phys. Rev. Lett.},\ }\textbf {\bibinfo {volume}
  {103}},\ \bibinfo {pages} {070503} (\bibinfo {year}
  {2009}{\natexlab{a}})}\BibitemShut {NoStop}%
\bibitem [{\citenamefont {Aulbach}(2011){\natexlab{a}}}]{aulbach2011symmetric}%
  \BibitemOpen
  \bibfield  {author} {\bibinfo {author} {\bibfnamefont {M.}~\bibnamefont
  {Aulbach}},\ }\href@noop {} {\bibfield  {journal} {\bibinfo  {journal} {Arxiv
  preprint arXiv:1103.0271}} (\bibinfo {year}
  {2011}{\natexlab{a}})}\BibitemShut {NoStop}%
\bibitem [{\citenamefont {Ribeiro}\ and\ \citenamefont
  {Mosseri}(2011)}]{PhysRevLett.106.180502}%
  \BibitemOpen
  \bibfield  {author} {\bibinfo {author} {\bibfnamefont {P.}~\bibnamefont
  {Ribeiro}}\ and\ \bibinfo {author} {\bibfnamefont {R.}~\bibnamefont
  {Mosseri}},\ }\Doi {10.1103/PhysRevLett.106.180502} {\bibfield  {journal}
  {\bibinfo  {journal} {Phys. Rev. Lett.},\ }\textbf {\bibinfo {volume}
  {106}},\ \bibinfo {pages} {180502} (\bibinfo {year} {2011})}\BibitemShut
  {NoStop}%
\bibitem [{\citenamefont {Markham}(2011)}]{PhysRevA.83.042332}%
  \BibitemOpen
  \bibfield  {author} {\bibinfo {author} {\bibfnamefont {D.~J.~H.}\
  \bibnamefont {Markham}},\ }\Doi {10.1103/PhysRevA.83.042332} {\bibfield
  {journal} {\bibinfo  {journal} {Phys. Rev. A},\ }\textbf {\bibinfo {volume}
  {83}},\ \bibinfo {pages} {042332} (\bibinfo {year} {2011})}\BibitemShut
  {NoStop}%
\bibitem [{\citenamefont {Aulbach}\ \emph {et~al.}(2010)\citenamefont
  {Aulbach}, \citenamefont {Markham},\ and\ \citenamefont
  {Murao}}]{aulbach2010maximally}%
  \BibitemOpen
  \bibfield  {author} {\bibinfo {author} {\bibfnamefont {M.}~\bibnamefont
  {Aulbach}}, \bibinfo {author} {\bibfnamefont {D.}~\bibnamefont {Markham}}, \
  and\ \bibinfo {author} {\bibfnamefont {M.}~\bibnamefont {Murao}},\
  }\href@noop {} {\bibfield  {journal} {\bibinfo  {journal} {New Journal of
  Physics},\ }\textbf {\bibinfo {volume} {12}},\ \bibinfo {pages} {073025}
  (\bibinfo {year} {2010})}\BibitemShut {NoStop}%
\bibitem [{\citenamefont {Horodecki}\ \emph {et~al.}(2009)\citenamefont
  {Horodecki}, \citenamefont {Horodecki}, \citenamefont {Horodecki},\ and\
  \citenamefont {Horodecki}}]{RevModPhys.81.865}%
  \BibitemOpen
  \bibfield  {author} {\bibinfo {author} {\bibfnamefont {R.}~\bibnamefont
  {Horodecki}}, \bibinfo {author} {\bibfnamefont {P.}~\bibnamefont
  {Horodecki}}, \bibinfo {author} {\bibfnamefont {M.}~\bibnamefont
  {Horodecki}}, \ and\ \bibinfo {author} {\bibfnamefont {K.}~\bibnamefont
  {Horodecki}},\ }\Doi {10.1103/RevModPhys.81.865} {\bibfield  {journal}
  {\bibinfo  {journal} {Rev. Mod. Phys.},\ }\textbf {\bibinfo {volume} {81}},\
  \bibinfo {pages} {865} (\bibinfo {year} {2009})}\BibitemShut {NoStop}%
\bibitem [{\citenamefont {Hardy}(1993)}]{PhysRevLett.71.1665}%
  \BibitemOpen
  \bibfield  {author} {\bibinfo {author} {\bibfnamefont {L.}~\bibnamefont
  {Hardy}},\ }\Doi {10.1103/PhysRevLett.71.1665} {\bibfield  {journal}
  {\bibinfo  {journal} {Phys. Rev. Lett.},\ }\textbf {\bibinfo {volume} {71}},\
  \bibinfo {pages} {1665} (\bibinfo {year} {1993})}\BibitemShut {NoStop}%
\bibitem [{\citenamefont {Hardy}(1994)}]{PhysRevLett.73.2279}%
  \BibitemOpen
  \bibfield  {author} {\bibinfo {author} {\bibfnamefont {L.}~\bibnamefont
  {Hardy}},\ }\Doi {10.1103/PhysRevLett.73.2279} {\bibfield  {journal}
  {\bibinfo  {journal} {Phys. Rev. Lett.},\ }\textbf {\bibinfo {volume} {73}},\
  \bibinfo {pages} {2279} (\bibinfo {year} {1994})}\BibitemShut {NoStop}%
\bibitem [{\citenamefont {Ghosh}\ and\ \citenamefont
  {Roy}(2010)}]{ghosh2010chainjournal}%
  \BibitemOpen
  \bibfield  {author} {\bibinfo {author} {\bibfnamefont {S.}~\bibnamefont
  {Ghosh}}\ and\ \bibinfo {author} {\bibfnamefont {S.}~\bibnamefont {Roy}},\
  }\href@noop {} {\bibfield  {journal} {\bibinfo  {journal} {Journal of
  Mathematical Physics},\ }\textbf {\bibinfo {volume} {51}},\ \bibinfo {pages}
  {1} (\bibinfo {year} {2010})}\BibitemShut {NoStop}%
\bibitem [{\citenamefont {Bell}\ \emph {et~al.}(1964)\citenamefont {Bell} \emph
  {et~al.}}]{bell1964einstein}%
  \BibitemOpen
  \bibfield  {author} {\bibinfo {author} {\bibfnamefont {J.}~\bibnamefont
  {Bell}} \emph {et~al.},\ }\href@noop {} {\bibfield  {journal} {\bibinfo
  {journal} {Physics},\ }\textbf {\bibinfo {volume} {1}},\ \bibinfo {pages}
  {195} (\bibinfo {year} {1964})}\BibitemShut {NoStop}%
\bibitem [{\citenamefont {Clauser}\ \emph {et~al.}(1969)\citenamefont
  {Clauser}, \citenamefont {Horne}, \citenamefont {Shimony},\ and\
  \citenamefont {Holt}}]{PhysRevLett.23.880}%
  \BibitemOpen
  \bibfield  {author} {\bibinfo {author} {\bibfnamefont {J.~F.}\ \bibnamefont
  {Clauser}}, \bibinfo {author} {\bibfnamefont {M.~A.}\ \bibnamefont {Horne}},
  \bibinfo {author} {\bibfnamefont {A.}~\bibnamefont {Shimony}}, \ and\
  \bibinfo {author} {\bibfnamefont {R.~A.}\ \bibnamefont {Holt}},\ }\Doi
  {10.1103/PhysRevLett.23.880} {\bibfield  {journal} {\bibinfo  {journal}
  {Phys. Rev. Lett.},\ }\textbf {\bibinfo {volume} {23}},\ \bibinfo {pages}
  {880} (\bibinfo {year} {1969})}\BibitemShut {NoStop}%
\bibitem [{app()}]{appendices}%
  \BibitemOpen
  \href@noop {} {\enquote {\bibinfo {title} {See supplemental material at [url
  will be inserted by publisher] for proof},}\ }\BibitemShut {NoStop}%
\bibitem [{\citenamefont {Penrose}(2000)}]{penrose20001}%
  \BibitemOpen
  \bibfield  {author} {\bibinfo {author} {\bibfnamefont {R.}~\bibnamefont
  {Penrose}},\ }\enquote {\bibinfo {title} {{On Bell non-locality without
  probabilities: some curious geometry}},}\ in\ \href@noop {} {\emph {\bibinfo
  {booktitle} {Quantum reflections}}}\ (\bibinfo  {publisher} {Cambridge
  University Press},\ \bibinfo {year} {2000})\ p.~\bibinfo {pages}
  {1}\BibitemShut {NoStop}%
\bibitem [{\citenamefont {Aulbach}(2011){\natexlab{b}}}]{martinthesis}%
  \BibitemOpen
  \bibfield  {author} {\bibinfo {author} {\bibfnamefont {M.}~\bibnamefont
  {Aulbach}},\ }\emph {\bibinfo {title} {Classification of Entanglement in
  Symmetric States}},\ \href@noop {} {Ph.D. thesis},\ \bibinfo  {school}
  {University of Leeds} (\bibinfo {year} {2011}{\natexlab{b}})\BibitemShut
  {NoStop}%
\bibitem [{Note1()}]{Note1}%
  \BibitemOpen
  \bibinfo {note} {Even though it would be useful in an LHV test to also test
  entanglement class in a device independent way, it does not make sense to
  assert that all parties `measure in the same basis'. One may append to the
  list of conditions (\ref {p1})-(\ref {p5}) some extra conditions which
  effectively imply the state is symmetric with respect to the basis given by
  measurement setting $1$. However, whilst one may be able to define Hardy type
  paradoxes which can be used to identify classes of entanglement in this way,
  these conditions would be difficult to fit into an inequality, and further it
  is not clear that it would be possible at all that such inequalities would
  also strictly separate degeneracy classes.}\BibitemShut {Stop}%
\bibitem [{\citenamefont {Gisin}(1991)}]{gisin1991bell}%
  \BibitemOpen
  \bibfield  {author} {\bibinfo {author} {\bibfnamefont {N.}~\bibnamefont
  {Gisin}},\ }\href@noop {} {\bibfield  {journal} {\bibinfo  {journal} {Physics
  Letters A},\ }\textbf {\bibinfo {volume} {154}},\ \bibinfo {pages} {201}
  (\bibinfo {year} {1991})}\BibitemShut {NoStop}%
\bibitem [{\citenamefont {\.{Z}ukowski}\ \emph {et~al.}(2002)\citenamefont
  {\.{Z}ukowski}, \citenamefont {Brukner}, \citenamefont {Laskowski},\ and\
  \citenamefont {Wie\'{s}niak}}]{PhysRevLett.88.210402}%
  \BibitemOpen
  \bibfield  {author} {\bibinfo {author} {\bibfnamefont {M.}~\bibnamefont
  {\.{Z}ukowski}}, \bibinfo {author} {\bibfnamefont {{\v C}.}~\bibnamefont
  {Brukner}}, \bibinfo {author} {\bibfnamefont {W.}~\bibnamefont {Laskowski}},
  \ and\ \bibinfo {author} {\bibfnamefont {M.}~\bibnamefont {Wie\'{s}niak}},\
  }\Doi {10.1103/PhysRevLett.88.210402} {\bibfield  {journal} {\bibinfo
  {journal} {Phys. Rev. Lett.},\ }\textbf {\bibinfo {volume} {88}},\ \bibinfo
  {pages} {210402} (\bibinfo {year} {2002})}\BibitemShut {NoStop}%
\bibitem [{\citenamefont {Prevedel}\ \emph {et~al.}(2009)\citenamefont
  {Prevedel}, \citenamefont {Cronenberg}, \citenamefont {Tame} \emph
  {et~al.}}]{PhysRevLett.103.020503}%
  \BibitemOpen
  \bibfield  {author} {\bibinfo {author} {\bibfnamefont {R.}~\bibnamefont
  {Prevedel}}, \bibinfo {author} {\bibfnamefont {G.}~\bibnamefont
  {Cronenberg}}, \bibinfo {author} {\bibfnamefont {M.~S.}\ \bibnamefont
  {Tame}},  \emph {et~al.},\ }\Doi {10.1103/PhysRevLett.103.020503} {\bibfield
  {journal} {\bibinfo  {journal} {Phys. Rev. Lett.},\ }\textbf {\bibinfo
  {volume} {103}},\ \bibinfo {pages} {020503} (\bibinfo {year}
  {2009})}\BibitemShut {NoStop}%
\bibitem [{\citenamefont {Wieczorek}\ \emph {et~al.}(2009)\citenamefont
  {Wieczorek}, \citenamefont {Krischek}, \citenamefont {Kiesel} \emph
  {et~al.}}]{PhysRevLett.103.020504}%
  \BibitemOpen
  \bibfield  {author} {\bibinfo {author} {\bibfnamefont {W.}~\bibnamefont
  {Wieczorek}}, \bibinfo {author} {\bibfnamefont {R.}~\bibnamefont {Krischek}},
  \bibinfo {author} {\bibfnamefont {N.}~\bibnamefont {Kiesel}},  \emph
  {et~al.},\ }\Doi {10.1103/PhysRevLett.103.020504} {\bibfield  {journal}
  {\bibinfo  {journal} {Phys. Rev. Lett.},\ }\textbf {\bibinfo {volume}
  {103}},\ \bibinfo {pages} {020504} (\bibinfo {year} {2009})}\BibitemShut
  {NoStop}%
\bibitem [{\citenamefont {Bastin}\ \emph
  {et~al.}(2009){\natexlab{b}}\citenamefont {Bastin}, \citenamefont {Thiel},
  \citenamefont {von Zanthier} \emph {et~al.}}]{PhysRevLett.102.053601}%
  \BibitemOpen
  \bibfield  {author} {\bibinfo {author} {\bibfnamefont {T.}~\bibnamefont
  {Bastin}}, \bibinfo {author} {\bibfnamefont {C.}~\bibnamefont {Thiel}},
  \bibinfo {author} {\bibfnamefont {J.}~\bibnamefont {von Zanthier}},  \emph
  {et~al.},\ }\Doi {10.1103/PhysRevLett.102.053601} {\bibfield  {journal}
  {\bibinfo  {journal} {Phys. Rev. Lett.},\ }\textbf {\bibinfo {volume}
  {102}},\ \bibinfo {pages} {053601} (\bibinfo {year}
  {2009}{\natexlab{b}})}\BibitemShut {NoStop}%
\end{thebibliography}%

\section{appendix}
\subsection{Proof of Proposition 1}\label{dickeproof}
Before proving the main proposition, we recall some properties of the Majorana representation~\cite{majorana1932atomi}~\cite{PhysRevA.83.042332}~\cite{aulbach2010maximally}~\cite{penrose20001} (For a comprehensive introduction, see~\cite{martinthesis}, Chapter 3).

Given a permutation symmetric state $\ket{\psi}$ of $n$ qubits, we can decompose it using the Majorana representation either as the sum of permutations of $n$ MPs (with possible duplication):
\begin{align}
\ket{\psi}=K\sum_{perm}\ket{\eta_1\ldots\eta_n}\label{decomp},
\end{align}
or as the sum of permutations of $l$ distinct MPs $|\eta_i\rangle$, each with degeneracy $d_i$:
\begin{align}
\ket{\psi}=K\sum_{perm}\ket{\eta_1^{d_1}\eta_2^{d_2}\ldots\eta_l^{d_l}},\label{multdecomp}\\
\forall i\neq j, \ket{\eta_i}\neq\ket{\eta_j}, \sum_{i=1}^l d_i=n.\nonumber
\end{align}

To find the MPs of $\ket{\psi}$, we use the coherent state decomposition. Let $\ket{\alpha}=\ket{0}+\alpha\ket{1}$, where $\alpha$ is a complex number. The overlap with $\ket{\psi}$ is
\begin{align}
P(\alpha)=(\bra{\alpha})^{\otimes n}\ket{\psi}\label{polynomial},
\end{align}
where $P(\alpha)$ is a complex polynomial with $\alpha$ as its indeterminate. The fundamental theorem of algebra states that~(\ref{polynomial}) has exactly $n$ roots, counting multiplicity (infinity is also considered a root if the degree of $P(\alpha)$ is smaller than $n$). These $n$ roots give $n$ points on the Riemann sphere via stereographic projection. Since $S^2\cong\mathbb{C}\mathbf{P}^1$, there are $n$ unordered points on the Bloch sphere corresponding to these $n$ roots. The MPs of $\ket{\psi}$ are given by the \emph{antipodal} points of the points representing the roots on the Bloch sphere. In the case where infinity is counted as a root, the corresponding MP would be $\ket{1}$. By definition, a qubit $\ket{\eta}$ is an MP of a state $\ket{\psi}$ of $n$ qubits if and only if
\begin{align}
(\bra{\eta^{\perp}})^{\otimes n}\ket{\psi}=0\label{orthogonal},
\end{align}
where $\ip{\eta^{\perp}}{\eta}=0$. For MPs with degeneracy $d_i$,~(\ref{orthogonal}) becomes
\begin{align}
(\bra{\eta_i^{\perp}})^{\otimes c}\ket{\psi}=0,\label{multi_ortho}
\end{align}
where $c\geq n-d_i+1$.

Now we prove an important lemma upon which the rest of the proof will rely. The purpose of this lemma is to show that orthogonality conditions like~(\ref{orthogonal}) and~(\ref{multi_ortho}) are only true if the state which we take the tensor product of (the \emph{bra} half) is an MP of the permutation symmetric state (the \emph{ket} half), and the order of tensor products depends on the degeneracy of the MP.

\begin{lemma}\label{lemma0}
If $\ket{\psi}$ is a permutation symmetric state of $n$ qubits with (distinct) MPs $\{\ket{\eta_1},\ket{\eta_2},\ldots,\ket{\eta_l}\}$, each having degeneracy $\{d_1,d_2,\ldots,d_l\}$, then \begin{equation}
(\bra{\theta})^{\otimes c}\ket{\psi}=0
\end{equation}
if and only if  $\ket{\theta}=\ket{\eta_i^{\perp}}$,  $\ip{\eta_i^{\perp}}{\eta_i}=0$ for some $i$, and $c\geq n-d_i+1$ (or equivalently, $d_i\geq n-c+1$).
\end{lemma}
\begin{proof}
The \emph{if} direction follows simply from expanding $\ket{\psi}$ using (\ref{decomp}),(\ref{multdecomp}) and the condition on MPs in the definition above. We will now focus on the \emph{only if} direction.

First notice that if $(\bra{\theta})^{\otimes c}\ket{\psi}=0$ with $c\leq n$, then $(\bra{\theta})^{\otimes n}\ket{\psi}=0$. As explained above, $(\bra{\theta})^{\otimes n}\ket{\psi}=0$ is only possible if $\ket{\theta}$ is an antipodal point of some MP of $\ket{\psi}$. Therefore $\ket{\theta}=\ket{\eta_i^{\perp}}$, $\ip{\eta_i^{\perp}}{\eta_i}=0$ for some $i$.

Now we want to show that $(\bra{\eta_i^{\perp}})^{\otimes c}\ket{\psi}=0$ implies $c\geq n-d_i+1$. Instead we will show the equivalent statement: $c< n-d_i+1$ (or equivalently $c\leq n-d_i$) implies $(\bra{\eta_i^{\perp}})^{\otimes c}\ket{\psi}\neq0$.

If $c=n-d_i$, then
\begin{align}
&(\bra{\eta_i^{\perp}})^{\otimes n-d_i}\ket{\psi}\label{l01}\\
&=(n-d_i)!(\bra{\eta_i^{\perp}})^{\otimes n-d_i}\ket{\underbrace{\eta_1\eta_2\ldots\eta_l}_{\{1\ldots l\}\setminus i}}(\ket{\eta_i})^{\otimes d_i}\neq 0.\label{l02}
\end{align}

To get from~(\ref{l01}) to~(\ref{l02}), we used that fact that when $c=n-d_i$, if we expand $\ket{\psi}$ using (\ref{decomp}),(\ref{multdecomp}) all other terms disappear. This term cannot be zero since by assumption no other MPs are equal to $|\eta_i\rangle$.

When $c<n-d_i$, if $(\bra{\eta_i^{\perp}})^{\otimes c}\ket{\psi}=0$, then this would imply $(\bra{\eta_i^{\perp}})^{\otimes n-d_i}\ket{\psi}=0$, which contradicts the argument just given. Thus when $c\leq n-d_i$, $(\bra{\eta_i^{\perp}})^{\otimes c}\ket{\psi}\neq0$.
\end{proof}

The statement of Proposition 1 refers to the state $\ket{\psi_i}=\ip{\eta_i}{\psi}$. Neither the statement nor this proof depend on the choice of $\ket{\eta_i}$. In fact what will be shown in this proof is that for any choice of $\ket{\eta_i}$, for the conditions in Proposition 1 to be satisfied, all other MPs of $\ket{\psi}$ are either orthogonal to it or lie on top of it, which is the definition of a (possibly rotated) Dicke state. For convenience, we now fix this $\ket{\eta_i}$ to be $\ket{\eta_1}$, so $\ket{\psi_i}$ will now be fixed to be $\ket{\psi_1}$.

The state $\ket{\psi}$ can be decomposed into MPs $\ket{\eta_i^{d_i}}$ as in~(\ref{multdecomp}). Similarly, the state $\ket{\psi_1}=\ip{\eta_1}{\psi}$ can be decomposed into
\begin{align}\label{multdef}
\ket{\psi_1}=K\sum_{perm}\ket{\mu_1^{m_1}\mu_2^{m_2}\ldots\mu_k^{m_k}},\\
\forall i\neq j, \ket{\mu_i}\neq\ket{\mu_j}, \sum_{i=1}^{k} m_i=n-1.
\end{align}

Apart from $\ket{\psi_1}$, there is another $(n-1)$-qubit permutation symmetric state useful to us, which is the state composed of all MPs of $\ket{\psi}$ except $\ket{\eta_i}$:
\begin{align}
\ket{\psi_{/i}}:=K\sum_{perm}\ket{\underbrace{\eta_1\ldots\eta_n}_{\{1\ldots n\}\setminus i}}.
\end{align}

By using Lemma~\ref{lemma0}, we can prove two corollaries about the states $\ket{\psi_1}$, $\ket{\psi_{/i}}$ and the degeneracies of their MPs:

\begin{corollary}\label{col1}
$(\bra{\gamma})^{\otimes c}\ket{\psi_1}=0$ if and only if $\ket{\gamma}=\ket{\mu_i^{\perp}}$, $\ip{\mu_i^{\perp}}{\mu_i}=0$ for some $i$ and $c\geq n-m_i$ (or equivalently, $m_i\geq n-c$).
\end{corollary}
\begin{proof}
Follows directly from Lemma~\ref{lemma0} by noticing that $\ket{\psi_1}$ is a permutation symmetric state of $n-1$ qubits.
\end{proof}

\begin{corollary}\label{col2}
$(\bra{\gamma})^{\otimes c}\ket{\psi_{/i}}=0$ if and only if $\ket{\gamma}=\ket{\eta_j^{\perp}}$, $\ip{\eta_j^{\perp}}{\eta_j}=0$ for some $j$, and 1). if $j\neq i$, then $c\geq n-d_j$ (or equivalently, $d_j\geq n-c$); 2). if $j=i$, then $d_j>1$ and $c\geq n-d_j+1$ (or equivalently, $d_j\geq n-c+1$).
\end{corollary}
\begin{proof}
When $j\neq i$, the proof follows directly from Lemma~\ref{lemma0}, in the same way as the proof of the previous corollary. When $j=i$, the degeneracy of $\ket{\eta_j}$ in $\ket{\psi_{/i}}$ is $d_j-1$, then by Lemma~\ref{lemma0} it can only be zero for $d_i>1$ and then $d_j-1\geq n-c\leftrightarrow d_j\geq n-c+1$.
\end{proof}

We now proceed to the three main lemmas of the proof. For clarity, we continue to use the notation that $\ket{\eta_i}$ is an MP of $\ket{\psi}$ with degeneracy $d_i$, and $\ket{\mu_i}$ is an MP of $\ket{\psi_1}$ with degeneracy $m_i$. We also use $K$ to denote the global normalization constant.

\begin{lemma}\label{lemma1}
$\forall i$, $\ket{\eta_i}=\ket{\mu_i}$, with $m_i\geq d_i-1.$
\end{lemma}
\begin{proof}
First of all, if $d_i=1$, then the statement is always true, so now we will focus on the case when $d_i>1$.

From Corollary~\ref{col1}, $(\bra{\eta_i^{\perp}})^{\otimes n-d_i+1}\ket{\psi_{1}}=0$ if and only if $\ket{\eta_i}=\ket{\mu_i}$ and $m_i\geq n-(n-d_i+1)=d_i-1$. So it suffices to show that $(\bra{\eta_i^{\perp}})^{\otimes n-d_i+1}\ket{\psi_{1}}=0$.

$\ket{\psi_1}$ can be decomposed into
\begin{align}
\ket{\psi_1}=K\sum_{l}\ip{\eta_1}{\eta_l}\ket{\psi_{/l}}.
\end{align}

Using this decomposition, we have
\begin{align}
&(\bra{\eta_i^{\perp}})^{\otimes n-d_i+1}\ket{\psi_{1}}\\
&=K(\bra{\eta_i^{\perp}})^{\otimes n-d_i+1}\sum_{l}\ip{\eta_1}{\eta_l}\ket{\psi_{/l}}\\
&=K\sum_{l}\ip{\eta_1}{\eta_l}(\bra{\eta_i^{\perp}})^{\otimes n-d_i+1}\ket{\psi_{/l}}\label{l13}\\
&=0,\label{l14}
\end{align}
where we used both cases in Corollary~\ref{col2} to get from (\ref{l13}) to (\ref{l14}).
\end{proof}

\begin{lemma}\label{lemma2}
$\forall i$, if $\ket{\eta_i}=\ket{\mu_i}$ and $m_i\geq d_i$, then $\ip{\eta_1}{\eta_i}=0$.
\end{lemma}
\begin{proof}
By Corollary~\ref{col1}, if $\ket{\eta_i}=\ket{\mu_i}$ and $m_i\geq d_i$, then $(\bra{\eta_i^{\perp}})^{\otimes n-d_i}\ket{\psi_{1}}=0$.

Using the same decomposition of $\ket{\psi_1}$ as in the proof of the previous lemma, we have:
\begin{align}
&(\bra{\eta_i^{\perp}})^{\otimes n-d_i}\ket{\psi_{1}}\\
&=K\sum_{j}\ip{\eta_1}{\eta_j}(\bra{\eta_i^{\perp}})^{\otimes n-d_i}\ket{\psi_{/j}}.
\end{align}

From Corollary~\ref{col2}, we can deduce that the only term which does not vanish in the sum is when $j=i$, thus we have:
\begin{align}
&(\bra{\eta_i^{\perp}})^{\otimes n-d_i}\ket{\psi_{1}}\\
&=K\ip{\eta_1}{\eta_i}\underbrace{(\bra{\eta_i^{\perp}})^{\otimes n-d_i}\ket{\psi_{/i}}}_{*}\label{l21}\\
&=0.
\end{align}

Since neither $K$ nor $*$ in (\ref{l21}) is $0$, we can conclude that $\ip{\eta_1}{\eta_i}=0$.

\end{proof}

\begin{lemma}\label{lemma3}
$\forall i$, if $\ket{\eta_i}=\ket{\mu_i}$ then $m_i\leq d_i$.
\end{lemma}
\begin{proof}
If $m_i>d_i$, then by Lemma~\ref{lemma2} $\ip{\eta_1}{\eta_i}=0$. We can write $\ket{\eta_i^{\perp}}$ as $\ket{\eta_1}$.

Partially expanding $\ket{\psi}$ in the $\{\ket{\eta_1},\ket{\eta_1^{\perp}}\}$ basis gives
\begin{align}
\ket{\psi}=K(\ket{\eta_1}\ket{\psi_1}+\ket{\eta_1^{\perp}}\ket{\widetilde{\psi_1}})
\end{align}

Now consider $(\bra{\eta_i^{\perp}})^{\otimes n-d_i}\ket{\psi}$:
\begin{align}
&(\bra{\eta_i^{\perp}})^{\otimes n-d_i}\ket{\psi}\\
&=(\bra{\eta_i^{\perp}})^{\otimes n-d_i}K(\ket{\eta_1}\ket{\psi_1}+\ket{\eta_1^{\perp}}\ket{\widetilde{\psi_1}})\label{l31}\\
&=K(\bra{\eta_i^{\perp}})^{\otimes n-d_i-1}\ket{\psi_1}\label{l32}\\
&=0\label{l33},
\end{align}
where we used the fact that $\ket{\eta_i^{\perp}}=\ket{\eta_1}$ to go from (\ref{l31}) to (\ref{l32}). We used Corollary~\ref{col1} and our assumption that $m_i>d_i$ to go from (\ref{l32}) to (\ref{l33}).

Clearly, the conclusion that $(\bra{\eta_i^{\perp}})^{\otimes n-d_i}\ket{\psi}=0$ contradicts Lemma~\ref{lemma0}, so $m_i\leq d_i$.
\end{proof}

Combining Lemmas~\ref{lemma1}, \ref{lemma2}, \ref{lemma3}, we get the main corollary which will prove the proposition.
\begin{corollary}\label{lemma4}
$\forall i$, $\ket{\eta_i}=\ket{\mu_i}$ with degenerecies such that either 1). $m_i=d_i$, $\ip{\eta_1}{\eta_i}=0$ or 2). $m_i=d_i-1$, $\ip{\eta_1}{\eta_i}\neq0$.
\end{corollary}

To finish the proof of Proposition 1, we first observe that since $\sum_{k} m_k=n-1$, and $n=\sum d_i$, we have
\begin{align}\label{eqn: sum di}
\sum_k m_k=n-1&=(\sum_k d_k)-1\\
&=\underbrace{(d_i-1)}_{\text{group 1}}+\underbrace{\sum_{j\neq i} d_j}_{\text{group 2}},
\end{align}
where we have separated the $d_i$ into two groups, group one, where the minus one is associated a particular $d_i$, and group two, the remaining $d_j$. Next we consider what is implied by the condition $S_{\psi_i} \subseteq S_\psi$ in Proposition 1 - i.e., where all Majorana points $|\mu_i\rangle$ coincide with $|\eta_i\rangle$. It is clear from Corollary~\ref{lemma4} and (\ref{eqn: sum di}) that this can only be done if group one is given by $d_1$, and group two comes from one Majorana point which is orthogonal to $|\eta_1\rangle$. Thus there are only two Majorana points of $|\psi\rangle$, $|\eta_1\rangle$ and $|\eta_2\rangle=|\eta_1^{\perp}\rangle$. This is exactly a Dicke state up to rotation of the Majorana sphere. Substitute any $\ket{\eta_i}$ for $\ket{\eta_1}$ in the beginning, the reasoning above is still valid.

The proof of this proposition, as shown in the main text, allows us to extend the Hardy paradox to $n$ parties and a constructive procedure to find the suitable measurement bases for all symmetric states except Dicke states. The fact that the procedure works for everything but Dicke states can be seen as a generalization to the fact that the original Hardy paradox works for all entangled states except the maximally entangled states. In the bipartite case, all states are LU-equivalent to a symmetric state, and the maximally entangled states are precisely bipartite Dicke states up to rotation. The reason for this, hinted by Hardy in~\cite{PhysRevLett.71.1665}, is that in order to have the paradox, we need some kind of ``symmetry breaking'', where after the projection on one MP, the symmetry of the remaining MPs will change. Dicke states, however, are ``too symmetric'' such that when projected on one of its MPs, the remaining MPs do not change.

\subsection{Proof of Proposition 2}\label{inequalityproof}

Assuming
\begin{align}
P(r_1,\ldots,r_n|M_1,\ldots,M_n)=\int \rho(\lambda)\prod_i P(r_i|M_i,\lambda) d\lambda\label{lhv},
\end{align}
 joint probabilities are products of probabilities of each party ( here we dropped $\lambda$ because we are always integrating over $d\lambda$). The Bell operator $\mathcal{P}^n$ can be rewritten as:
\begin{align}
&P_1(0)P_2(0)\ldots P_n(0)\nonumber\\
-&P_1(1)P_2(1)\ldots P_n(1)\nonumber\\
-&(1-P_1(1))P_2(0)\ldots P_n(0)\nonumber\\
&\vdots\nonumber\\
-&P_1(0)\ldots P_{n-1}(0)(1-P_n(1)),\label{lhs}
\end{align}
by noting
\begin{align*}
P_i(0)=P_i(0|0),\\
P_i(1)=P_i(1|1).
\end{align*}
The subscripts denote different parties and essentially show that probabilities of obtaining the same result by different parties are independent.
An expansion gives:
\begin{align}
P_1(0)P_2(0)\ldots P_n(0)&-P_1(1)P_2(1)\ldots P_n(1)\label{row_1}\\
+P_1(1)P_2(0)\ldots P_n(0)&-P_2(0)P_3(0)\ldots P_n(0)\label{row_2}\\
+P_1(0)P_2(1)\ldots P_n(0)&-P_1(0)P_3(0)\ldots P_n(0)\label{row_3}\\
\vdots\nonumber\\
+P_1(0)P_2(0)\ldots P_n(1)&-P_1(0)P_2(0)\ldots P_{n-1}(0)\label{row_4}
\end{align}
Note that the rows~(\ref{row_2}) to~(\ref{row_4}) are all less than or equal to $0$, because for all $0\leq P_i\leq 1$,
\begin{align}\label{negative_prop}
\prod_{i=1}^{n}P_i\leq\prod_{i=1}^{n-1}P_i.
\end{align}
Factoring the second term in~(\ref{row_1}) and the first term in ~(\ref{row_2}), we have:
\begin{align}\label{negative_row}
P_1(1)(P_2(0)\ldots P_n(0)-P_2(1)\ldots P_n(1)).
\end{align}
If $P_2(0)\ldots P_n(0)\leq P_2(1)\ldots P_n(1)$, then~$(\ref{negative_row})\leq 0$ (by using (\ref{negative_prop}) on the first term in (\ref{row_1}) and the second term in (\ref{row_2}), and on rows (\ref{row_3}) to (\ref{row_4})), and there is nothing more to prove. Otherwise
\begin{align}
&P_1(1)(P_2(0)\ldots P_n(0)-P_2(1)\ldots P_n(1))\nonumber\\ \leq &(P_2(0)\ldots P_n(0)-P_2(1)\ldots P_n(1)),
\end{align}
which means $(\ref{lhs})$ is smaller than or equals to
\begin{align}
P_1(0)P_2(0)\ldots P_n(0)&-P_2(1)P_3(1)\ldots P_n(1)\\
+P_1(0)P_2(1)\ldots P_n(0)&-P_1(0)P_3(0)\ldots P_n(0)\\
\vdots\nonumber\\
+P_1(0)P_2(0)\ldots P_n(1)&-P_1(0)P_2(0)\ldots P_{n-1}(0).
\end{align}
Repeat the same procedure $n$ times, each time assuming $\prod_k P_k(0)>\prod_k P_k(1)$ (otherwise we can terminate the proof). What has been shown after these $n$ steps is $(\ref{lhs})\leq P_1(0)P_2(0)\ldots P_n(0)-P_1(0)P_2(0)\ldots P_{n-1}(0)$, whose right hand side is less than or equal to $0$ by~(\ref{negative_prop}).

\subsection{Proof of Proposition 3}\label{alldickeproof}
First we will write the inequality in Proposition 3 as a function of $n$, $k$ and $\theta$.
\begin{align}
&P(0\ldots0|0\ldots0)\nonumber\\
&=|\ip{+,\ldots ,+}{S(n,k)}|^2\nonumber\\
&=((\frac{1}{\sqrt{2}})^n {n\choose k}^{-\frac{1}{2}} {n\choose k})^2,\label{d1}\\
&P(0\ldots0|1\ldots0)\nonumber\\
&=|(\cos{\frac{\theta}{2}}\bra{0}-\sin{\frac{\theta}{2}}\bra{1})\otimes\ip{+,\ldots ,+}{S(n,k)}|^2\nonumber\\
&=((\frac{1}{\sqrt{2}})^{n-1} {n\choose k}^{-\frac{1}{2}} (\cos{\frac{\theta}{2}} {n-1 \choose k}-\sin{\frac{\theta}{2}} {n-1 \choose k-1}))^2,\label{d2}\\
&P(1\ldots1|1\ldots1)\nonumber\\
&=|^{n\otimes}(\sin{\frac{\theta}{2}}\bra{0}+\cos{\frac{\theta}{2}}\bra{1})\ket{S(n,k)}|^2\nonumber\\
&=({n\choose k}^{-\frac{1}{2}} {n \choose k} (\cos{\frac{\theta}{2}})^k (\sin{\frac{\theta}{2}})^{n-k})^2.\label{d3}
\end{align}

To simplify our calculations, we divide each probability (\ref{d1}) - (\ref{d3}) by $n\choose k$. Doing this rescales the Bell operator in Proposition 2, which will not change its positivity property.
\begin{align}
&P'(0\ldots0|0\ldots0)\nonumber\\
&=(\frac{1}{2})^n,\label{scaled1}\\
&P'(0\ldots0|1\ldots0)\nonumber\\
&=(\frac{1}{2})^{n-1} {n\choose k}^{-2} (\cos{\frac{\theta}{2}} {n-1 \choose k}-\sin{\frac{\theta}{2}} {n-1 \choose k-1})^2,\label{scaled2}\\
&P'(1\ldots1|1\ldots1)\nonumber\\
&=(\cos{\frac{\theta}{2}})^{2k}(\sin{\frac{\theta}{2}})^{2n-2k}.\label{scaled3}
\end{align}

Because of the permutation symmetry of $\ket{S(n,k)}$, the $n$ probabilities $P'(0\ldots0|1\ldots0)$ to $P'(0\ldots0|0\ldots1)$ are all equal. After some simplification, the rescaled Bell operator becomes:
\begin{align}
P'(n,k,\theta)&=(\frac{1}{2})^n\nonumber\\
&-n(\frac{1}{2})^{n-1}(\frac{n-k}{n}\cos{\frac{\theta}{2}}-\frac{k}{n}\sin{\frac{\theta}{2}})^2\nonumber\\
&-(\cos{\frac{\theta}{2}})^{2k}(\sin{\frac{\theta}{2}})^{2n-2k}.\label{func}
\end{align}

First we note some properties of (\ref{scaled2}) and (\ref{scaled3}). For (\ref{scaled2}), it can always reach 0 for all $n, k$ when $\tan{\frac{\theta}{2}}=\frac{n-k}{k}$. (\ref{scaled3}) is 0 when $\theta=0$ and $\theta=\pi$, and it reaches its maximum when $(\cos{\frac{\theta}{2}})^{2k}=(\sin{\frac{\theta}{2}})^{2n-2k}$. Also, if we fix $\theta$ and $n$, its derivative with respect to $k$ is 
\begin{align}
2(\cos{\frac{\theta}{2}})^{2k}(\sin{\frac{\theta}{2}})^{2n-2k}(\log{\cos{\frac{\theta}{2}}}-\log{\sin{\frac{\theta}{2}}}),
\end{align}
which means that for fixed $\theta$ and $n$, when $\theta<\frac{\pi}{2}$, (\ref{scaled3}) is monotonically increasing with respect to $k$, when $\theta>\frac{\pi}{2}$, (\ref{scaled3}) is monotonically decreasing with respect to $k$, and when $\theta=\frac{\pi}{2}$, (\ref{scaled3}) is independent of $k$.

Now consider the equation $n\times(\ref{scaled2})=(\frac{1}{2})^{n+1}$:
\begin{align}
n(\frac{1}{2})^{n-1}(\frac{n-k}{n}\cos{\frac{\theta}{2}}-\frac{k}{n}\sin{\frac{\theta}{2}})^2=(\frac{1}{2})^{n+1},\\
\implies (\frac{n-k}{n}\cos{\frac{\theta}{2}}-\frac{k}{n}\sin{\frac{\theta}{2}})^2=\frac{1}{4n}.\label{equation}
\end{align}

After taking the square root of both sides, moving the term with $\sin{\frac{\theta}{2}}$ to one side, substuting $\sin{\frac{\theta}{2}}$ with $\sqrt{1-\cos^2{\frac{\theta}{2}}}$ and square both sides again, (\ref{equation}) can be seen as a quadratic equation having $\cos{\frac{\theta}{2}}$ as unknown. It may have 0, 1 or 2 roots when $\theta$ takes any value in the interval $[ 0, \pi]$. If it has no root, then we know $n\times(\ref{scaled2})< (\frac{1}{2})^{n+1}$ for all values of $\theta$ (because (\ref{scaled2}) can always reach 0). Also note that (\ref{scaled3}) is zero when $\theta=0$ and $\theta=\pi$, so if (\ref{equation}) has no root then $(\ref{func})>0$ when $\theta=0$ and $\theta=\pi$. Similarly, if (\ref{equation}) has one root, we can show that $(\ref{func})>0$ when $\theta=0$ or $\theta=\pi$, depending on the location of the root: when the root is smaller than $\frac{\pi}{2}$, we take $\theta=\pi$, otherwise we take $\theta=0$.

We now proceed to the case when (\ref{equation}) has two roots. By solving the quadratic equation we can have closed forms of the two roots in $[ 0, \pi]$, which we call $\theta_+$ and $\theta_-$:
\begin{align}
\cos{\frac{\theta_+}{2}}&=\frac{k\sqrt{n}-n\sqrt{n}+\sqrt{8k^4-k^2n-8k^3n+4k^2n^2}}{2(2k^2-2kn+n^2)}\label{r+}\\
\cos{\frac{\theta_-}{2}}&=\frac{k\sqrt{n}-n\sqrt{n}-\sqrt{8k^4-k^2n-8k^3n+4k^2n^2}}{2(2k^2-2kn+n^2)}\label{r-}
\end{align}

What we want to show now is that for any fixed $n$, $(\cos{\frac{\theta_+}{2}})^{2k}(\sin{\frac{\theta_+}{2}})^{2n-2k}< (\frac{1}{2})^{n+1}$ and $(\cos{\frac{\theta_-}{2}})^{2k}(\sin{\frac{\theta_-}{2}})^{2n-2k}< (\frac{1}{2})^{n+1}$ for all $k$. From the monotonicity of (\ref{scaled3}), it suffices to show that if the inequalities hold for $k=\frac{n}{2}$, then they must hold for all $k$. So for now we will only consider the case when $n$ is even. We also restrict ourselves to the positive root. The argument below is symmetric, with appropriate sign/monotonicity changes it also applies to the negative root. 

When $n$ is even and $k=\frac{n}{2}$, $(\cos{\frac{\theta_+}{2}})^{2k}(\sin{\frac{\theta_+}{2}})^{2n-2k}$ simplify to:
\begin{align}
(\frac{1}{32})^k\underbrace{(\frac{-\sqrt{k}+\sqrt{k(4k-1)}}{k})^{2k}}_{*}\underbrace{(2+\frac{\sqrt{(4k-1)}}{k})^k}_{**}.\label{2k}
\end{align}

To get the upper bound on (\ref{2k}), we first rewrite $(*)$ and $(**)$:
\begin{align}
(*)&=((\frac{-\sqrt{k}+\sqrt{k(4k-1)}}{k})^2)^k=(4-\frac{2\sqrt{4k-1}}{k})^k\\
&=(4-\frac{4\sqrt{k-\frac{1}{4}}}{k})^k=(4(1-\frac{\sqrt{k-\frac{1}{4}}}{k}))^k,\label{b1}\\
(**)&=(2+\frac{2\sqrt{k-\frac{1}{4}}}{k})^k=(2(1+\frac{\sqrt{k-\frac{1}{4}}}{k}))^k.\label{b2}
\end{align}

Substitute $(\ref{b1})$ and $(\ref{b2})$ into (\ref{2k}), we get
\begin{align}
(\frac{1}{32}\times 4\times 2\times (1-\frac{k-\frac{1}{4}}{k^2}))^k=(\frac{1}{2})^{2k}(1-\frac{k-\frac{1}{4}}{k^2})^k.\label{bound}
\end{align}

Note that $\lim_{n\to \infty} (1-\frac{1}{x})^x=e^{-1}$, and it approaches this limit from below. We thus have 
\begin{align}
(1-\frac{k-\frac{1}{4}}{k^2})^k< (1-\frac{1}{k})^k\leq e^{-1}<\frac{1}{2}\label{exp}
\end{align}

From (\ref{bound}) and (\ref{exp}), we get the desired bound
\begin{align}
&(\cos{\frac{\theta_+}{2}})^{2k}(\sin{\frac{\theta_+}{2}})^{2n-2k}\nonumber\\
&< (\frac{1}{2})^{2k}(1-\frac{1}{k})^k<(\frac{1}{2})^{2k+1}
\end{align}

To show that the proposition holds for odd $n$, we let $n=2k+1$ for some $k$, then (\ref{func}) becomes
\begin{align}
&(\frac{1}{2})^{2k+1}-(2k+1)(\frac{1}{2})^{2k}(\frac{k+1}{2k+1}\cos{\frac{\theta}{2}}-\frac{k}{2k+1}\sin{\frac{\theta}{2}})^2\\
-&(\cos{\frac{\theta}{2}})^{2k}(\sin{\frac{\theta}{2}})^{2k+2}\\
=&\frac{1}{2}((\frac{1}{2})^{2k}\\
-&(2k+1)(\frac{1}{2})^{2k-1}(\frac{k+1}{2k+1}\cos{\frac{\theta}{2}}-\frac{k}{2k+1}\sin{\frac{\theta}{2}})^2\label{odd1}\\
-&2(\sin{\frac{\theta}{2}})^2(\cos{\frac{\theta}{2}})^{2k}(\sin{\frac{\theta}{2}})^{2k}\label{odd2}
)
\end{align}

It is easy to see that $\frac{k+1}{2k+1}\geq\frac{1}{2}$ and $\frac{k}{2k+1}\leq\frac{1}{2}$, so (\ref{odd1}) can be bounded by
\begin{align}
(\ref{odd1})\geq 2k(\frac{1}{2})^{2k-1}(\frac{1}{2}\cos{\frac{\theta}{2}}-\frac{1}{2}\sin{\frac{\theta}{2}})^2.\label{oddsub1}
\end{align}

Assume $\cos{\frac{\theta_+}{2}}\leq\frac{1}{\sqrt{2}}$ (the other case will be discussed below), which means $2(\sin{\frac{\theta}{2}})^2\geq 1$, we can bound (\ref{odd2}) by:
\begin{align}
(\ref{odd2})\geq (\cos{\frac{\theta}{2}})^{2k}(\sin{\frac{\theta}{2}})^{2k}.\label{oddsub2}
\end{align}

Substitute (\ref{oddsub1}) and (\ref{oddsub2}) into (\ref{odd1}) and (\ref{odd2}), we have 
\begin{align}
&(\frac{1}{2})^{2k+1}\nonumber\\
&-(2k+1)(\frac{1}{2})^{2k}(\frac{k+1}{2k+1}\cos{\frac{\theta}{2}}-\frac{k}{2k+1}\sin{\frac{\theta}{2}})^2\nonumber\\
&-(\cos{\frac{\theta}{2}})^{2k}(\sin{\frac{\theta}{2}})^{2k+2}\nonumber\\
\leq&\frac{1}{2}((\frac{1}{2})^{2k}\nonumber\\
&-2k(\frac{1}{2})^{2k-1}(\frac{1}{2}\cos{\frac{\theta}{2}}-\frac{1}{2}\sin{\frac{\theta}{2}})^2\nonumber\\
&-(\cos{\frac{\theta}{2}})^{2k}(\sin{\frac{\theta}{2}})^{2k}\nonumber
),
\end{align}
which is just two times the expression for when $n=2k$. If $\cos{\frac{\theta_+}{2}}\leq\frac{1}{\sqrt{2}}$, we can let $n=2k-1$, and get a similar argument. For the negative root, the the same reasoning follows.

To sum up, for even $n$, if we let $n\times(\ref{scaled2})=(\frac{1}{2})^{n+1}$, then $(\ref{scaled3})<(\frac{1}{2})^{n+1}$, which means $(\ref{func})>0$. For odd $n$, depending on the value of $\cos{\frac{\theta_+}{2}}$ or $\cos{\frac{\theta_-}{2}}$, we can always bound (\ref{func}) by considering the closest even $n$. So (\ref{func}) can always be positive for some value of $\theta$.

\end{document}